\documentclass[twocolumn]{aastex631}

\usepackage{multirow}

\usepackage{CJKutf8}
\newcommand{\zh}[1]{\begin{CJK}{UTF8}{gbsn}#1\end{CJK}}

\begin{document}

\title{PEPSI Investigation, Retrieval, and Atlas of Numerous Giant Atmospheres (PIRANGA). IV. High-Resolution Phased-Resolved Spectroscopy of The Ultra Hot Jupiter KELT-20 b}

\author[0000-0001-5330-7709]{Victoria Bonidie}
\affiliation{Department of Physics and Astronomy, University of Pittsburgh, 3941 O'Hara Street, Pittsburgh, PA 15260, USA}
\affiliation{Pittsburgh Particle Physics, Astrophysics, and Cosmology Center (PITT PACC), University of Pittsburgh, Pittsburgh, PA 15260, USA}

\author[0000-0002-5099-8185]{Marshall C. Johnson}
\affiliation{Department of Astronomy, The Ohio State University, 4055 McPherson Laboratory, 140 West 18th Avenue, Columbus, OH 43210 USA}
\author[0000-0002-4361-8885]{Ji Wang \zh{(王吉)}}
\affiliation{Department of Astronomy, The Ohio State University, 4055 McPherson Laboratory, 140 West 18th Avenue, Columbus, OH 43210 USA}
\author[0009-0008-8016-6591]{Sydney Petz}
\affiliation{Department of Astronomy, The Ohio State University, 4055 McPherson Laboratory, 140 West 18th Avenue, Columbus, OH 43210 USA}
\affiliation{Department of Physics and Astronomy, Michigan State University, East Lansing, MI 48824, USA}
\author{Jake Kamen}
\affiliation{Department of Astronomy, The Ohio State University, 4055 McPherson Laboratory, 140 West 18th Avenue, Columbus, OH 43210 USA}
\author[0009-0001-1459-3738]{Calder Lenhart}
\affiliation{Department of Astronomy, The Ohio State University, 4055 McPherson Laboratory, 140 West 18th Avenue, Columbus, OH 43210 USA}
\author[0000-0002-4531-6899]{Alison Duck}
\affiliation{Department of Astronomy, The Ohio State University, 4055 McPherson Laboratory, 140 West 18th Avenue, Columbus, OH 43210 USA}
\affiliation{Jet Propulsion Laboratory, California Institute of Technology, Pasadena, CA 91109, USA}
\author[0000-0003-3494-343X]{Carles Badenes}
\affiliation{Department of Physics and Astronomy, University of Pittsburgh, 3941 O'Hara Street, Pittsburgh, PA 15260, USA}
\affiliation{Pittsburgh Particle Physics, Astrophysics, and Cosmology Center (PITT PACC), University of Pittsburgh, Pittsburgh, PA 15260, USA}
\author[0000-0002-6192-6494]{Klaus Strassmeier}
\affiliation{Leibniz-Institute for Astrophysics Potsdam (AIP), An der Sternwarte 16, D-14482 Potsdam, Germany}
\affiliation{Institute of Physics \& Astronomy, University of Potsdam, Karl-Liebknecht-Str. 24/25, D-14476 Potsdam, Germany}
\author[0000-0002-0551-046X]{Ilya Ilyin}
\affiliation{Leibniz-Institute for Astrophysics Potsdam (AIP), An der Sternwarte 16, D-14482 Potsdam, Germany}

\begin{abstract}
We present five datasets of high-resolution optical emission spectra of the ultra-hot Jupiter KELT-20 b with the PEPSI spectrograph. Using a Bayesian retrieval framework, we constrain its dayside pressure-temperature profile and abundances of Fe, Ni, and Ca, providing the first measurements for Ni and Ca for KELT-20 b in emission. We retrieve the pre- and post-eclipse datasets separately (corresponding to the evening and morning sides, respectively), and compare the constraints on their thermal structures and chemical abundances. We constrain lower abundances in the pre-eclipse datasets compared to the post-eclipse datasets. We interpret these results with an equilibrium chemistry model which suggests $\sim 10-30 \times$ supersolar refractory abundances. Due to the well-known degeneracy between absolute abundances and continuum opacities, the abundance ratios are more precise probes of the planetary abundances. Therefore we measure the abundance ratios [Ni/Fe] and [Ca/Fe] across these datasets and find they agree within 1$\sigma$. We constrain [Ni/Fe] to be consistent with solar within 2$\sigma$, and [Ca/Fe] to be 0.001-0.01$\times$ solar, not accounting for ionization. We compare these abundance ratios with literature results for KELT-20 b in transmission, and find they agree within 2$\sigma$, suggesting that even though the abundances vary significantly as a function of phase, the abundance ratios of these species remain relatively constant. We find a $\sim100$ K difference in temperature at the top of the thermal inversion, suggesting a hotter evening side than morning side and underscoring the importance of considering 3D effects when studying ultra-hot Jupiters.
\end{abstract}

\section{Introduction}
Since the discovery of the first hot Jupiter exoplanet by \cite{Mayor+Queloz}, the field of Jupiter-like exoplanets has evolved rapidly, now entering a phase where we can answer questions about their atmospheric composition and temperature structure. Advancements in both ground-based spectrographs with resolving powers exceeding $R = 25,000$, as well as improved techniques in data processing methodology, have allowed us to access the hundreds to thousands of individual absorption and emission lines from the planetary atmosphere that would be unresolved at lower spectral resolution.

High-resolution cross-correlation spectroscopy (HRCCS) has emerged as a powerful tool to enable precise detections of the chemical composition of close-in planets\citep[e.g.][]{Snellen+10, Brogi+12}. This technique takes advantage of the fact that planetary spectral lines are Doppler shifted during an observation, carrying the planetary spectrum across multiple resolution elements of the spectrograph, whereas the spectral contaminants (telluric and stellar) remain quasi-stationary, allowing the planetary signal to be disentangled.

Furthermore, applying these methods on multi-phase datasets have enabled us to begin studying the 3D structure of these planets. HRS observations in transmission and phase-curves have detected species through cross-correlation across different planetary phases \citep[e.g.][]{Lenhart+25, Mraz+24, Wardenier+24, Ehrenreich+20}. These studies have targeted ultra-hot Jupiters (UHJs), a class of gas giant exoplanets with equilibrium temperatures exceeding 2200 K. These extreme temperatures increase their scale heights, improving their transmission detectability, and increases their flux ratio with respect to the star, improving their detectability for emission spectroscopy, making them ideal targets in HRCCS studies. UHJs are expected to be tidally locked, therefore performing HRCCS studies as a function of orbital phase probes different planetary longitudes.

While cross-correlation is an efficient method to detect chemical species in the atmospheres of exoplanets, it does not provide quantitative constraints on their atmospheric properties. To overcome this limitation, novel techniques have been developed to ``map'' the cross-correlation to a likelihood \citep{BrogiLine19, Gibson+20}, which can then be integrated into a Bayesian framework to explore the range of atmospheric properties that fit a given spectrum. This technique, known as an atmospheric retrieval, has been widely adopted in the field and has demonstrated the ability to obtain quantitative constraints on atmospheric properties, such as pressure-temperature (P-T) profiles and chemical abundances. Atmospheric retrievals of UHJs are often applied to datasets with limited phase coverage or lower resolution spectra, where multi-epoch spectral information cannot be resolved, effectively reducing their atmospheric properties to one dimension. However, there has been new focus on performing atmospheric retrievals as a function of planetary phases in both transmission \citep[e.g][]{Maguire+24, Gandhi+22}, and in emission \citep[e.g][]{Ramkumar+25, vanSluijs+23, Pino+22}. This new frontier of high-resolutions multi-phase observations of UHJ targets opens new opportunities to place quantitative atmospheric constraints at different planetary longitudes. 

In this paper, we present a phase-resolved Bayesian atmospheric retrieval analysis of KELT-20 b,\citep{Lund+17}, also known as MASCARA-2 b \citep{Talens+18}, an UHJ orbiting a bright A2-type star with an orbital period of $\sim$3.5 days. It has been the target of many high-resolution spectroscopic studies in both emission and transmission, yielding detections of several refractory metals such as Fe, Na, Mg, Mn, Ni, Cr, Ca, Si, Ti, and V \citep[e.g.][]{Casasayas-Barris+19, Stangret+20, Hoeijmakers+2020, Nugroho+20, Bello-Arufe+2022, borsa+22, cont+22, Yan+22, Gandhi+23, Johnson+23, Petz+24}
as well as molecular species such as H$_2$O and CO \citep[e.g.][]{Finnerty+25, Kasper+23} in its atmosphere. We perform separate retrievals on the combined, pre-eclipse, and post-eclipse datasets to place quantitative constraints on its vertical temperature structure and chemical abundances and explore the difference between the morning and evening sides. 

This paper is organized as follows: in Section \ref{sec:obs}, we present our observations and data preparation; in Section \ref{sec:retrieval}, we outline our retrieval framework; in Section \ref{sec:results} we present and contextualize our retrieval results for each dataset by comparing them to literature values and equilibrium chemistry models before concluding in Section \ref{sec:conclusion}.

\section{Observations and Data Preparation}\label{sec:obs}
\begin{figure}
\centering
\includegraphics[width=\columnwidth]{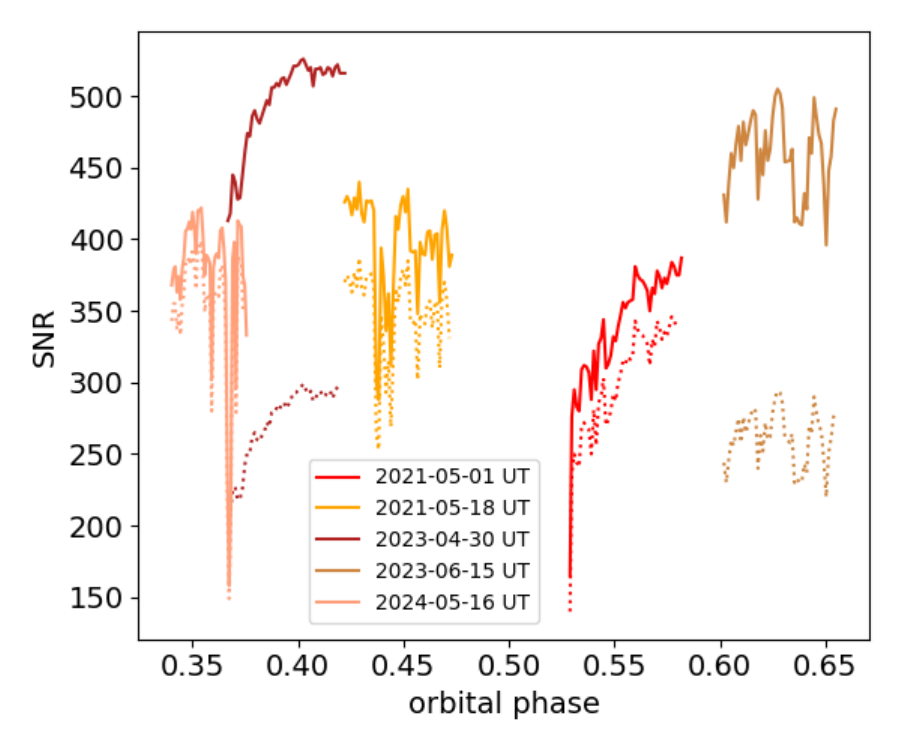}
\caption{Signal-to-noise ratio of our observations as a function of orbital phase. Each observation was taken with a 300 second exposure time. The PEPSI red and blue arms are shown as the solid and dotted lines, respectively. The SNR values reported correspond to the 95th percentile per-pixel signal-to-noise ratios for each spectrum. The change from CD II to CD III in the 2023 observations resulted in a significant decrease in SNR in the blue arm.}

\label{fig:phases}
\end{figure}

We obtained our data on KELT-20 b using the Potsdam Échelle Polarimetric and Spectrographic Instrument (PEPSI; \cite{Strassmeier+15}) on the Large Binocular Telescope (LBT) located on Mt. Graham, Arizona, USA. We use the time-series spectra of the pre- (2021-05-18 UT) and post- (2021-05-01 UT) eclipse observations of KELT-20 b, presented in \cite{Johnson+23}, along with three additional observations (two pre-eclipse and one post-eclipse) further from secondary eclipse. See Table \ref{observations}, and Figure \ref{fig:phases} for further information of these observations. 

\begin{figure}
\centering
\includegraphics[width=0.9\columnwidth]{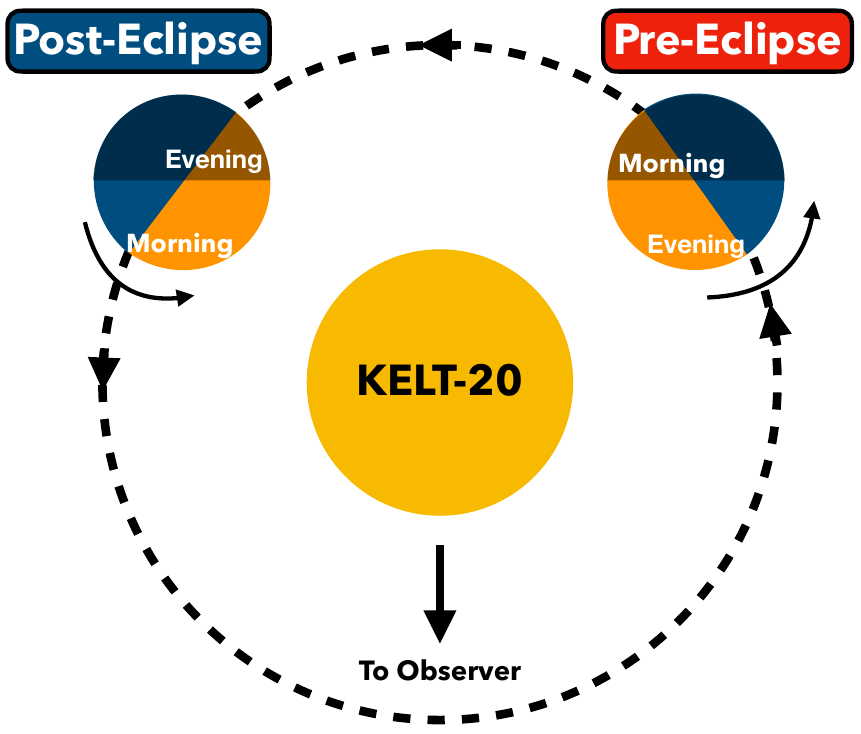}
\caption{Top-down view of KELT-20 b's orbit as it passes behind its host star, moving counterclockwise from pre-eclipse to post-eclipse. The orange side depicts the hot day side, while the blue side corresponds to the cooler night side of the tidally-locked planet. The translucent gray side depicts the side of the planet that is outside of the view of the observer, who is positioned off the bottom of the page.}

\label{fig:cartoon}
\end{figure}

As illustrated in Figure \ref{fig:cartoon}, the pre-eclipse datasets observe more of the evening side of KELT-20 b, while the post-eclipse datasets probe the morning side. All observations were taken using the 200 $\micron$ fiber, which provides a constant resolving power of R = 130,000. For the observations in in 2021 and 2024, we used the PEPSI cross dispersers (CD) III and V, providing a wavelength coverage of 4800-5441 \r{A} and 6278-7419 \r{A} in the blue and red channels, respectively.  Further details of the 2021 observations can be found in \cite{Johnson+23} and \cite{Petz+24}.

For the observations in 2023, we switched to CD II instead of CD III due to a mechanical issue with the spectrograph, so we instead had the wavelength coverage of 4265–4800 \r{A} in the blue arm. The data were reduced using the \texttt{SDS4PEPSI} pipeline \citep{Ilyin2000, Strassmeier+18}, which performs a standard data reduction and outputs a 1-dimensional wavelength-calibrated, continuum normalized spectrum. The pipeline also estimates the variance in each pixel, which we use to propagate uncertainties throughout our analysis.

\begin{deluxetable*}{cccccccc}
\tablecaption{Log of Observations\label{observations}}
\tablewidth{0pt}
\tablehead{
Date (UT) & $N_{\mathrm{spec, blue}}$ & $N_{\mathrm{spec, red}}$ & Exp. Time (s) & Airmass Range & Phases Covered & SNR$_{\mathrm{blue}}$ & SNR$_{\mathrm{red}}$
}
\startdata
2021 May 1 & 46 & 47 & 300 &  1.01-2.03 & 0.53-0.58 & 301 & 340 \\
2021 May 18 & 44 & 45 & 300 & 1.00-1.48 & 0.42-0.47 & 347 & 397 \\
2023 April 30 & 45 & 45 & 300 & 1.00-2.10 & 0.37-0.42 & 272 & 498 \\
2023 June 15 & 46 & 46 & 300 & 1.00-1.17 & 0.60-0.66 & 244 & 431 \\
2024 May 16 & 42 & 44 & 200 & 1.00-1.17 & 0.34-0.37 & 351 & 374\\
\enddata
\tablecomments{$N_{\mathrm{spec}}$ is the number of spectra obtained on that night. Exp. time is the exposure time in seconds. SNR$_{\mathrm{blue}}$ and SNR$_{\mathrm{red}}$ are the nightly average of the $95^{\mathrm{th}}$ quantile per-pixel signal-to-noise ratios in the blue and red arms, respectively. In cases where the number of spectra does not match between the two arms, blue arm spectra in twilight have been excluded due to larger scattered sunlight contamination which is not significant in the red arm.} 
\end{deluxetable*}

\subsection{Extraction of the Planetary Spectra}
To extract the faint exoplanetary signal from contaminants, we take advantage of the fact that stellar and telluric lines remain (quasi-)static during the observation, while the spectrum of the planet undergoes a Doppler shift of the order of several tens of km s$^{-1}$. Our procedure mirrors that of \cite{Johnson+23}, and we provide only a brief summary here.
First, we run the \texttt{MOLECFIT} package \citep{Kausch+15, Smette+15} on the red arm of each PEPSI spectrum to model out the telluric lines. This treatment was not necessary for the blue arm as it was largely telluric free. Next, we stack the spectra to create a median-combined spectrum and subtract it from each of the time series spectra in order to remove stellar lines and time invariant components. 

We then use the SYSREM \citep{Tamuz+05} algorithm to correct for any residuals left over after the \texttt{MOLECFIT} treatment. We adapt the python implementation of SYSREM, \texttt{PySysRem}\footnote{https://github.com/stephtdouglas/PySysRem}, in order to apply it to the PEPSI data. The algorithm iteratively subtracts linear trends in time from each spectral pixel, thus removing the quasi-static telluric and stellar lines that were not fully removed by \texttt{MOLECFIT}, or any instrumental systematics, leaving residuals consisting only of noise and the planetary signal. 

In order to maximize the signal strength, we adopt the methodology presented in E. F. Spring \& J. L. Birkby (2025, in preparation) to stop SYSREM once the removal of a successive systematic fails to improve to residuals by more than 1 part in $10^{4}$. The  number of systematics removed is low, ranging from 0-2 per night and per arm.

\section{Retrieval Framework}\label{sec:retrieval}
\subsection{Forward Model}\label{Forward Model}
\begin{figure*}[t]
\centering
\includegraphics[width=\textwidth]{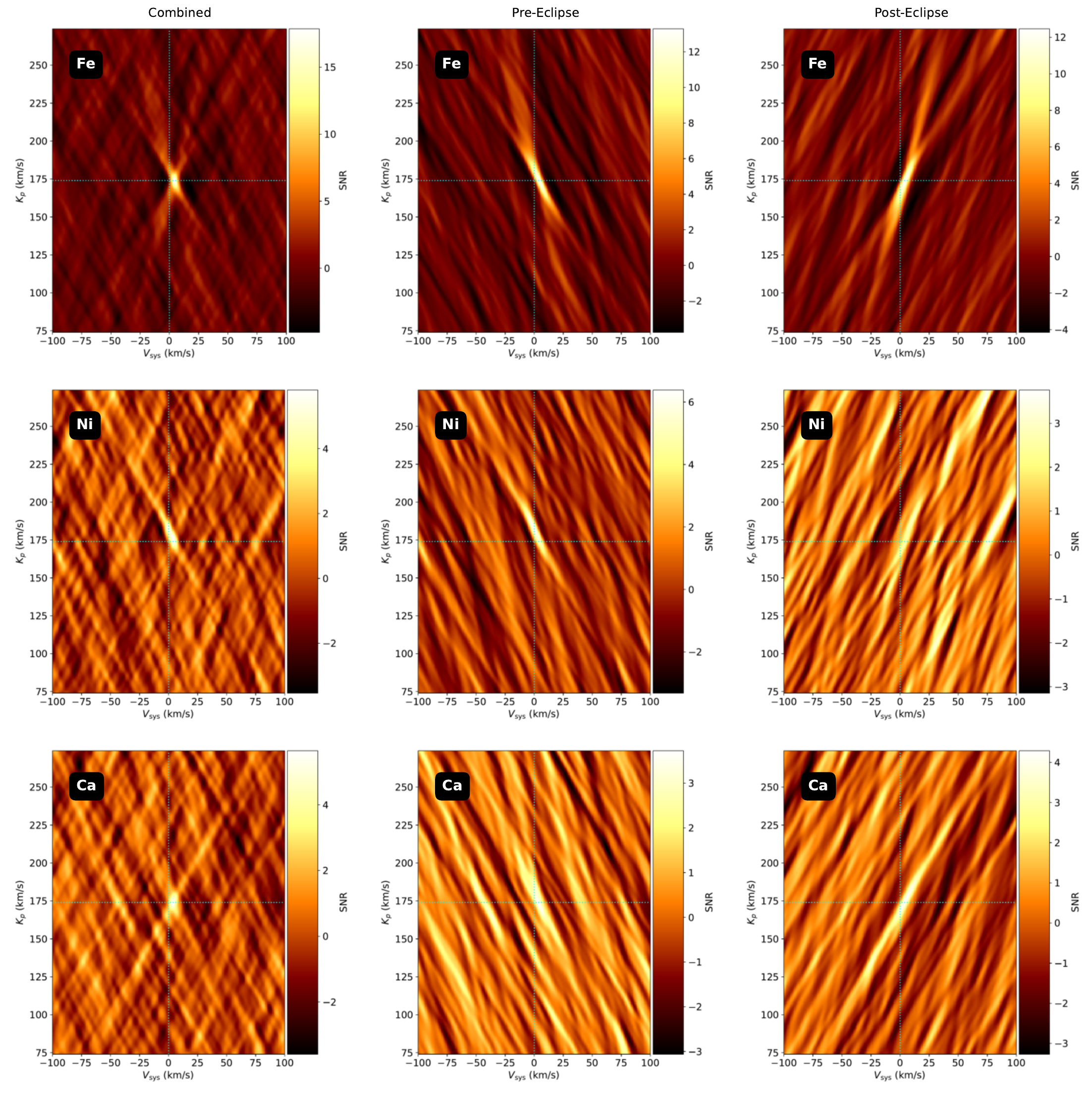}
\caption{Top row: shifted and combined CCF Fe, for the combined (left), pre-eclipse (middle) and post-eclipse (right) datasets. Middle row: same as the top row, but for Ni. Bottom row: same as the top row, but for Ca. Each $\geq3 \sigma$ detection falls within our tentative detection range. The vertical and horizontal dashed lines in all CCF maps represent the $K_p$ and $v_{sys}$ parameters for which we should expect to find a signal.}
\label{fig:ccfs}
\end{figure*}

We used the radiative transfer code \texttt{petitRADTRANS} \cite{molliere+19} at the default resolution ($R = 10^6$) to forward model the emission spectrum of KELT-20 b. We include continuum opacity collision-induced absorption of H$_2-$H$_2$ and H$_2-$He, and Rayleigh scattering due to H$_2$ and He, and H$^{-}$ bound-free and free-free absorption. We compute the volume mixing ratios (VMRs) for H$^-$, H, and e$^-$ using \texttt{FastChem} \citep{Stock+18} for each temperature structure produced at every iteration of the Monte-Carlo-Markov-Chain MCMC. This allows the continuum opacity species to vary without increasing the number of free parameters in our retrieval. We include the VMRs for Fe, Ni, and Ca as free parameters, assumed to be constant as a function of vertical depth. We chose these species as they each yield $\geq$ 4$\sigma$ detections in their combined cross-correlation functions (CCFs) (\citet{Petz+24}, Kamen et al. in prep), as well as greater than $\geq$ 3$\sigma$ detections in their pre- and post-eclipse CCFs, as shown in Figure \ref{fig:ccfs}.

We model the atmosphere of KELT-20 b using 100 log-uniform spaced pressure layers between $10^2$ and $10^{-6}$ bar and assume an inverted pressure–temperature (P--T) profile of the form given by Equation (29) \cite{Guillot+10} with four variables: $T_{irr}$ (irradiation temperature), $\kappa$ (infrared opacity), $\gamma$ (visible-to-infrared opacity ratio), and $T_{int}$ (intrinsic temperature). Of these variables, we fix $T_{int}$ to a constant 100 K, as is often done in other high resolution studies \citep[e.g.][]{Maguire+24, Ramkumar+25}, and leave the rest as free parameters in our retrieval.  
We assume a tidally locked rotation rate \citep[e.g.][]{Finnerty+25, Johnson+23} and convolve the models with a rotational broadening kernel using Equation (18.14) of \cite{Gray}, and an instrumental broadening kernel using a Gaussian line spread function with a width corresponding to the PEPSI resolving power of R = 130,000 \citep{Strassmeier+18}.
We then calculate the planet-to-star flux ratio by dividing the model by a Phoenix stellar model spectrum \citep{Husser+13} corresponding to the stellar effective temperature from \cite{Lund+17}.

The forward model spectra are then Doppler-shifted by linear interpolation to the radial velocity of each of our time series spectra given by:
\begin{equation}
    V_p = K_p \sin(2\pi\phi) + V_{sys},
\end{equation}

\noindent where $K_p$ is the planetary radial velocity semi-amplitude, $V_{sys}$ is the systemic velocity, and $\phi$ is the orbital phase of the planet.

\begin{deluxetable}{cccc}
\tabletypesize{\footnotesize}
\tablecaption{A summary of the stellar, planetary, and ephemeris values used
in our retrieval.\label{planetparams}}
\tablehead{
Parameter & Symbol (Unit) & Value & Source}
\startdata
Planet radius & $R_p\;(R_J)$ & 1.741 & a \\
Planet mass & $M_p\;(M_J)$ & 3.382 & a \\
Stellar radius & $R_*\;(R_\odot)$ & 1.565 & a \\
Stellar mass & $M_*\;(M_\odot)$ & 1.76 & a \\
Effective temperature & $T_{\text{eff}} \;(K)$ & 8720 & a \\
Systemic velocity & $V_{sys}$ (km s$^{-1}$) & 22.78 
& b \\
Orbital period & $P \;(d)$ & 3.47410151 
& c \\
Epoch of mid-transit & T$_0$ (BJD$_{\text{TDB}}$) & 2459757.811176 
& c \\
Transit duration & T$_{14}$ (d) & 0.147565 
& c
\enddata
\tablerefs{(a) \cite{Lund+17}; (b) \cite{Petz+24}; (c) \cite{Lenhart+25}} 
\end{deluxetable}

\subsection{Sysrem Distortion}
Applying SYSREM not only removes the static stellar and telluric signals in the data, but also distorts the underlying planetary spectrum. This effect must be accounted for in order to retrieve accurate parameters from the planetary spectra. We follow the methodology of \cite{Gibson+22} to apply a corresponding distortion to the model spectra. The corrected model is, from Eqn.~7 of \cite{Gibson+22}
\begin{equation}
    \mathbf{M}' = \mathbf{U}(\Lambda\mathbf{U})^{\dagger}(\Lambda\mathbf{M})
\end{equation}
where $\mathbf{M}$ is a matrix holding the Doppler-shifted forward model spectra for one night and one PEPSI arm, repeated for each night-arm combination; $\mathbf{U}$ is a matrix containing the SYSREM correction coefficients, with one additional row filled with ones in order to account for the effects of median-correcting the spectra; and $\Lambda$ is a diagonal matrix where the diagonal terms are the inverse of the mean variance of each spectrum. $^{\dagger}$ denotes the Moore-Penrose inverse, $\mathbf{X}^{\dagger}=(\mathbf{X}^\mathrm{T}\mathbf{X})^{-1}\mathbf{X}^\mathrm{T}$. We applied the filter to each model during the retrieval.

\subsection{Log Likelihood Analysis}
In order to compute the likelihood of the model fit to our data, we rely on the commonly used technique in high-resolution Bayesian statistics \citep{BrogiLine19, Gibson+20} to ``map'' cross-correlation values of an atmospheric model onto a log-likelihood value.
For this work, we compute the log-likelihood as defined in \cite{Gibson+20}, namely,
 \begin{equation}
     \ln \mathcal{L} = -\frac{N}{2}\ln 2\pi - N \ln \beta -\sum_i^N\ln\sigma_i-\frac{1}{2}\chi^2
 \end{equation}
 where 
 \begin{equation}
 \chi^2 = \frac{1}{\beta^2}\bigg(\sum_i^N \frac{f_i^2}{\sigma_i^2} + \alpha^2 \sum_i^N \frac{m_i^2}{\sigma_i^2} - 2 \alpha \sum_i^N \frac{f_i m_i}{\sigma_i^2}\bigg)
 \label{eqn:chisq}
 \end{equation}
 and where, for sums over pixels $i=1,\ldots,N$, $f_i$ and $\sigma_i$ are the flux and uncertainty in each pixel, $m_i$ is the model at each pixel (where each $m_i$ corresponds to a single element in a matrix $\mathbf{M}'$), and $\alpha$ and $\beta$ are multiplicative scaling parameters for the model and uncertainties, respectively. For simplicity, here we assume $\alpha = \beta=1$. Unlike \cite{Gibson+20}, we keep the constant terms because we compute $\ln\mathcal{L}$ separately for the blue and red arms of PEPSI as they contain a different number of spectra and must ensure that they are properly scaled when we combine them. The log-likelihood function is computed individually for each dataset, and these values are subsequently summed to derive the combined log-likelihood function.

\begin{deluxetable*}{lccccc} 
\tablecaption{Description of retrieved parameters (including prior ranges), derived parameters, and fixed parameters.\label{tab:parameters}}
\tablewidth{\textwidth}
\tablehead{\colhead{Parameter Name} & \colhead{Symbol} & \colhead{Prior} & \multicolumn{3}{c}{Value} \\ \cline{4-6} & & & \colhead{Pre-eclipse} & \colhead{Post-eclipse} & \colhead{Combined}}
\startdata
\multicolumn{6}{c}{\textit{Free Parameters}} \\ 
\hline
log Fe volume-mixing ratio & Fe & $\mathcal{U}(-12, -2)$ &$-4.29_{-0.15}^{+0.19}$ & $-3.49_{-0.38}^{+0.38}$ & $-3.96_{-0.17}^{+0.33}$ \\ 
log Ni volume-mixing ratio & Ni & $\mathcal{U}(-12, -2)$, &$-5.20_{-0.22}^{+0.24}$ & $-5.06_{-0.49}^{+0.43}$ & $-5.12_{-0.23}^{+0.34}$ \\ 
log Ca volume-mixing ratio & Ca & $\mathcal{U}(-12, -2)$ &$-7.83_{-0.30}^{+0.32}$ & $-6.94_{-0.43}^{+0.45}$ & $-7.46_{-0.24}^{+0.32}$ \\ 
log infrared opacity (cm$^2$ g$^{-1}$) & $\log(\kappa)$ & $\mathcal{U}(-4, 0)$ &$-1.14_{-0.15}^{+0.25}$ & $-0.56_{-0.41}^{+0.36}$ & $-0.89_{-0.22}^{+0.39}$ \\ 
log visible-to-infrared opacity & $\log(\gamma)$ & $\mathcal{U}(0, 2)$ &$1.01_{-0.05}^{+0.05}$ & $0.89_{-0.04}^{+0.05}$ & $0.94_{-0.04}^{+0.04}$ \\ 
Irradiation temperature (K) & $T_{\mathrm{irr}}$ & $\mathcal{G}(2862, 24)$ &$2863_{-24}^{+23}$ & $2862_{-25}^{+24}$ & $2864_{-24}^{+23}$ \\ 
RV offset (km s$^{-1}$) & $\Delta RV$ & $\mathcal{U}(-10,20)$ &$4.53_{-0.43}^{+0.45}$ & $5.65_{-0.46}^{+0.39}$ & $5.17_{-0.12}^{+0.12}$ \\ 
RV semi-amplitude (km s$^{-1}$) & $K_p$ & $\mathcal{U}(150,200)$ &$175.19_{-0.76}^{+0.72}$ & $174.89_{-0.84}^{+0.75}$ & $174.10_{-0.19}^{+0.21}$ \\  
\hline
\multicolumn{6}{c}{\textit{Derived Parameters}} \\ 
\hline
Nickel-to-Iron abundance ratio & [Ni/Fe]& ... &$0.11_{-0.29}^{+0.46}$ &$0.37_{-0.29}^{+0.31}$ &$-0.29_{-0.68}^{+0.60}$ \\ 
Calcium-to-Iron abundance ratio &[Ca/Fe] & ... &$-2.34_{-0.29}^{+0.45}$ &$-2.38_{-0.36}^{+0.38}$ &$-2.29_{-0.60}^{+0.61}$ \\
\enddata
\end{deluxetable*}

\subsection{Temperature Prior}\label{sec:Tirr}
To validate our retrieval framework, we tested it using mock data generated with the methods described in \S\ref{Forward Model}, but with the addition of Gaussian noise at a SNR of 400, representative of realistic observational conditions. We initially apply uniform or log-uniform priors for all parameters of interest, and use the \texttt{emcee} tool \citep{ForemanMackey+13} to compute their posteriors. The retrieval tightly constrained the VMRs, but weakly constrained irradiation temperature $T_{irr}$, resulting in a large dynamic range in the location of the inversion layer. Runs of our retrieval with no prior on $T_{\mathrm{irr}}$ resulted in temperatures ranging from 1400 to 3400 K, a far larger range than physically plausible. All of these solutions have a similar upper atmospheric temperature, as the emission-line data is primarily sensitive to the properties of the atmosphere at and above the inversion, and only poorly constrains the lower atmosphere. To address this limitation, we repeated the analysis with a Gaussian prior applied to $T_{irr}$. 
The resulting posteriors were consistent with those \textit{without} the addition of the temperature prior, but the resulting P-T profile had a more realistic spread of temperatures at the bottom of the inversion, demonstrating that a tighter prior on $T_{irr}$ is essential for our retrieval framework to effectively constrain the P-T profile when working with comparable datasets.

\begin{figure}
\centering
\includegraphics[width=0.9\columnwidth]{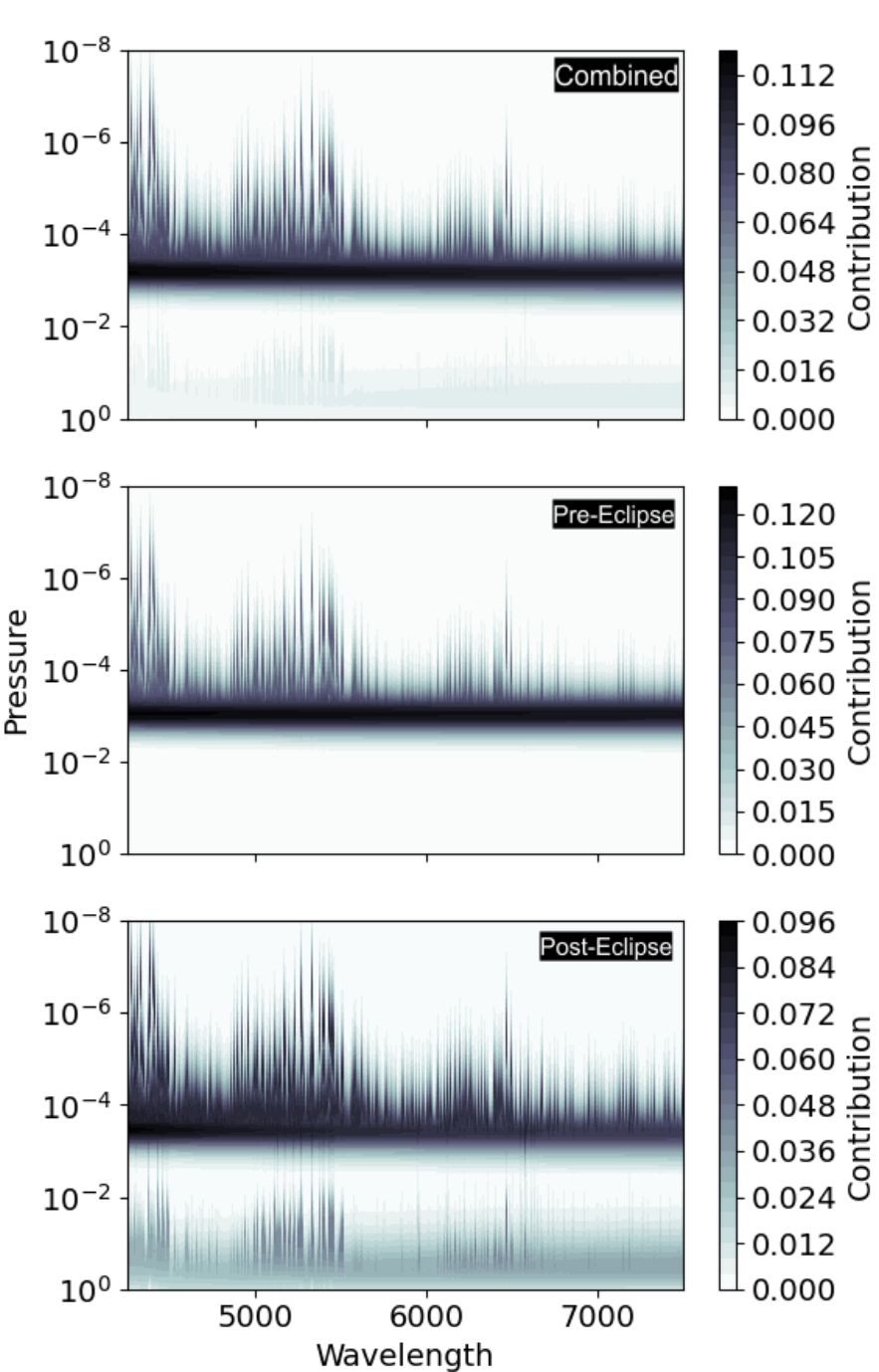}
\caption{Combined contribution function for Fe, Ni, and Ca given the best fit P-T profile from the combined, pre-eclipse, and post-eclipse retrievals. We note that the VMRs of Fe, Ni, and Ca are assumed to be constant as a function of pressure, while the continuum opacity species computed with FastChem (H, H$^{-}$, e$^{-}$) vary with altitude.}

\label{fig:contribution}
\end{figure}

In the \cite{Guillot+10} parameterization of the P-T profile, the quantity $T_{\mathrm{irr}}$ is the irradiation temperature at the sub-stellar point and is larger than the more commonly computed equilibrium temperature $T_{\mathrm{eq}}$ by a factor of $\sqrt{2}$. Empirically, for our case of a strong temperature inversion, $T_{\mathrm{irr}}$ is the temperature of the atmosphere near the bottom of the inversion (which is approximately isothermal at deeper pressures). Our emission line data has little to no constraining power on this parameter, as the emission lines are by definition formed at the inversion. We therefore choose to put a prior on $T_{\mathrm{irr}}$ as an  estimate of the planet's day-side temperature based on the atmospheric temperature measured from the TESS secondary eclipse. The TESS measurement is complementary to the high-resolution emission spectra as it provides information about the continuum that the emission spectra cannot probe, as shown in the contribution function in Figure \ref{fig:contribution}. The additional information from TESS helps to anchor our retrieval where our data does not probe.

The TESS \citep{TESS} 120 second pre-search data conditioning simple aperture photometry from sectors 40, 41, 54, 74  and 75 were detrended using a 3 times transit duration median filter implemented by \texttt{wotan} \citep{wotan}. Then \texttt{Exofastv2} \citep{exofastv2} was used to fit the primary and secondary eclipses of KELT-20. We measured a secondary eclipse depth of 140 $\pm$ 7 ppm. The secondary eclipse is then related to the planetary day-side temperature through Equation \ref{eqn:eclipsedepth} where $\delta_{sec}$ is the eclipse depth, $A_{\rm g}$ is the geometric albedo, a is the orbital semi-major axis, and $B_{\lambda}$  represents the Planck function integrated over the wavelengths of the TESS band-pass accounting for the band-pass response function. 

\begin{equation}
\label{eqn:eclipsedepth}
   \delta_{sec} = (\frac{R_{p}}{R_{*}})^2 \frac{B_{\lambda}(T_{p})}{B_{\lambda}(T_{*})} + A_{\rm g}(\frac{R_{p}}{a})^2
\end{equation}

\cite{Dang+25} shows compelling evidence for very low levels of albedo for highly irradiated planets so we fix the geometric albedo term to zero. Then, we employ an MCMC to sample ranges of planetary day-side temperatures which best reproduce our measured eclipse depth. Thus, we estimate the day-side temperature of KELT-20 b to be 2862 $\pm$ 24 K based on the TESS secondary eclipse depth. Similarly, \cite{Dang+25} estimates a day-side temperature for KELT-20b of 2820 $\pm$ 80 K based on Spitzer \citep{spitzer} 4.5$\mu$m phase curve observations.

\subsection{MCMC Set Up}
We apply the retrieval framework to the pre-eclipse, post-eclipse, and combined datasets, following the same steps as detailed in our mock retrieval analysis. For each retrieval, we run our MCMC with 40 walkers for 3,000 steps, with a burn-in length of 1,000, resulting in 80,000 samples of the posterior. All of the prior ranges are listed in Table \ref{tab:parameters}. We apply uniform or log-uniform priors on each of our free parameters, except for $T_{irr}$, for which we apply a Gaussian prior with the mean and standard deviation obtained from the methods described in \ref{sec:Tirr}.

\section{Results and Discussion}\label{sec:results}
\subsection{Retrieved Abundances}
\label{sec:abundances}
With all five nights of observation combined, we constrain the log of the VMRs of 
Fe ($-3.96^{+0.32}_{-0.17}$), 
Ni ($-5.13^{+0.33}_{-0.24}$), 
and Ca ($-7.46^{+0.32}_{-0.24}$). 
We note that this work provides the first constraints on Ni and Ca in emission for KELT-20 b. In our pre-eclipse only retrieval, we find slightly lower abundances of 
Fe ($-4.29^{+0.19}_{-0.15}$), 
Ni ($-5.20^{+0.24}_{-0.22}$), 
and Ca ($-7.83^{+0.32}_{-0.30}$) 
than the results from the combined datasets. In contrast, our post-eclipse retrieval constrains slightly higher abundances of 
Fe ($-3.49^{+0.38}_{-0.38}$), 
Ni ($-5.06^{+0.45}_{-0.43}$), 
and Ca ($-6.94^{+0.45}_{-0.43}$),
but all species' abundances remain consistent with the combined datasets within 1$\sigma$ due to larger uncertainties. For ease of comparison, all abundance constraints are listed side-by-side as log(VMRs) in Table \ref{tab:parameters} and their posteriors are shown in Figure \ref{fig:combined} for the combined datasets, and Figure \ref{fig:phase} for the pre- and post-eclipse datasets.

Our Fe abundances from each dataset are in agreement with those reported in \citet{Kasper+23}, 
but are $\sim$1-2 dex higher than those reported in \cite{Finnerty+25} in emission and \cite{Gandhi+23} in transmission. For Ni and Ca, we compare our values to those reported in \cite{Gandhi+23} with transmission spectra (listed as MASCARA-2 b). We find our Ni abundances exceed their upper limit of -6.04, and our Ca abundances exceed their constraint of $8.49^{+0.33}_{-0.31}$ by $\sim$1-2 dex.

The discrepancies in our retrieved abundances may suggest a longitudinally asymmetric distribution of metals across the surface of KELT-20 b. The pre-eclipse datasets are dominated by light from the evening side, where the atmospheric material is rotating from the hot day side into the cooler night side, whereas the post-eclipse datasets probe the morning side, receiving material from the cooler night side. Therefore, a depletion in the Fe and Ca abundances in the pre-eclipse data could be due to a higher ionization fraction on the evening side. However, we note that do not find significant ($>3\sigma$) evidence for these ionized species in the CCFs. While this can be explained for Ni+ and Ca+ which have few lines within the PEPSI bandpass, it is more puzzling for Fe+ which has many lines. When we inject Fe+ into our spectra at a VMR of $10^{-5}$ we detect it in the combined CCF with a $>10 \sigma$ significance, suggesting that Fe+ should be detectable if present within the pressure levels probed with PEPSI. Therefore, this simple assumption of ionization may not be capturing the mechanisms causing these discrepancies. Furthermore, if there were a significant difference in the ionization of refractory metals between the evening and morning sides, we would not expect the abundance of neutral Ni to stay relatively constant between both datasets.

\subsection{Comparison to FastChem}

\begin{figure*}[t]
\centering
\includegraphics[width=\textwidth]{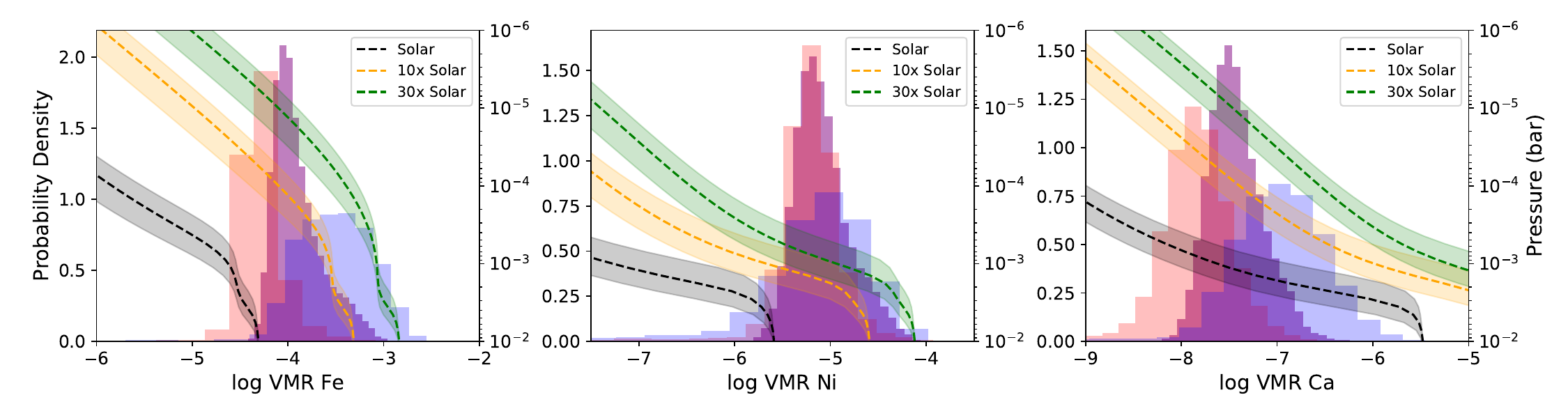}
\caption{Comparison of retrieved abundances to those predicted from FastChem equilibrium chemistry models. Each panel shows the histograms from the retrieved posteriors for the log VMRs of Fe, Ni, and Ca, respectively, from our pre-eclipse (red), post-eclipse (blue), and combined (purple) datasets. The dashed lines show the vertical abundances profiles from FastChem computed from our best fit P-T profile from our combined sample, and the shaded regions show that from our pre- and post-eclipse P-T profiles. The FastChem models assuming solar abundances are provided in black, the 10$\times$ in orange, and the 30$\times$ solar in green. Our retrieved abundances suggest a super-solar composition.}

\label{fig:fastchem}
\end{figure*}

In order to contextualize our results, we compare our retrieved abundances to FastChem equilibrium chemistry models, assuming solar, 10$\times$ solar, and 30$\times$ solar compositions. Figure \ref{fig:fastchem} shows the histograms of our retrieved posteriors compared to the computed FastChem models for Fe, Ni, and Ca, plotted as a function of pressure (i.e. altitude). As our contribution function in Figure \ref{fig:contribution} shows, we are able to probe layers between $10^{-3} - 10^{-6}$ bar, we find our retrieved abundances for Fe to be most consistent with a 10-30$\times$ solar model, whereas Ca is broadly consistent with both solar and super-solar models, depending on the exact pressure level probed.  However, Ni seems to only be consistent with the 30$\times$ solar model at the deepest pressure layers probed. Our assumed constant-with-altitude VMRs are more heavily weighted toward the deeper pressure layers before strong ionization occurs \citep[e.g.][]{Kasper+23}, so comparing our retrieved abundances to the lowest pressure levels probed is a justifiable comparison. However, we caution against relying on absolute abundance constraints alone to infer a metallicity, as we will discuss in \S\ref{sec:ratios} that high-resolution studies are better at constraining abundance ratios than absolute abundances.

\subsection{Abundance Ratios}\label{sec:ratios}
As previous works have demonstrated that high resolutions spectroscopy is more sensitive to the relative strengths of spectral lines than the absolute level \citep[e.g.][]{Gandhi+23, Gibson+22, Maguire+24}, we also measure the nickel-to-iron [Ni/Fe] and calcium-to-iron [Ca/Fe] ratios normalized to solar values calculated in \cite{Asplund+09}. This helps to mitigate the degeneracies between our retrieved abundances and the opacity level, i.e., high abundances with low opacity can produce the same emission features as low abundances with high opacity. This is especially true with high opacity species like Fe, so normalizing to Fe abundance helps constrain the relative concentrations independently of the opacity. We note that we do not correct for ionization effects when computing the abundance ratios, so they should be interpreted as lower limits.

\begin{figure}
\centering
\includegraphics[width=\columnwidth]{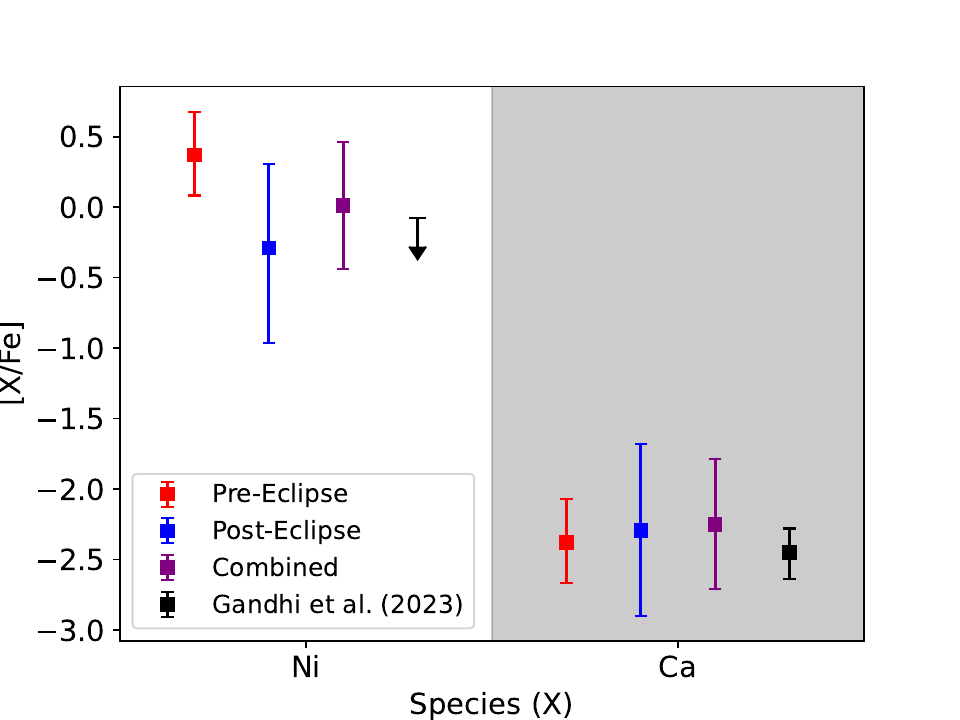}
\caption{Computed nickel-to-iron and calcium-to-iron abundance rations normalized to solar for the pre-eclipse (red), post-eclipse (blue), and combined (purple) datasets. Also plotted are the [Ni/Fe] and [Ca/Fe] abundance ratios from \cite{Gandhi+23}, computed for KELT-20 b from their constraints on Fe, Ni, and Ca from transmission spectroscopy.}
\label{fig:ratios}
\end{figure}

We measure a [Ni/Fe] ratio that is consistent with solar in the combined and post-eclipse datasets, but a slightly super-solar [Ni/Fe] in the pre-eclipse dataset. For [Ca/Fe], we measure a significantly subsolar (2-3 dex) abundance for each dataset, as listed in Table \ref{tab:parameters}. These constraints are broadly consistent with previous measurements for KELT-20 b in transmission in \cite{Gandhi+23}. We plot these values along with those from  \cite{Gandhi+23}  in Figure \ref{fig:ratios}.

Despite the differences in our retrieved posteriors for our VMRs, our calculated [Ni/Fe] and [Ca/Fe] ratios from the pre-, post-, and combined datasets are within 1$\sigma$ agreement with each other. Combined with the abundance ratios calculated by \cite{Gandhi+23}, this suggests that the [Ni/Fe] and [Ca/Fe] ratios stay relatively constant (within 2$\sigma$) throughout the secondary eclipse, as well as during transit. The consistency in these results across multiple observations, coupled with previous results using different instruments and retrieval frameworks, provides another testament of the reliability of high-resolution atmospheric retrieval \citep{Ramkumar+25, Maguire+24}.

\subsection{Pressure-Temperature Profile}
\begin{figure*}[t]
\centering
\includegraphics[width=0.7\textwidth]{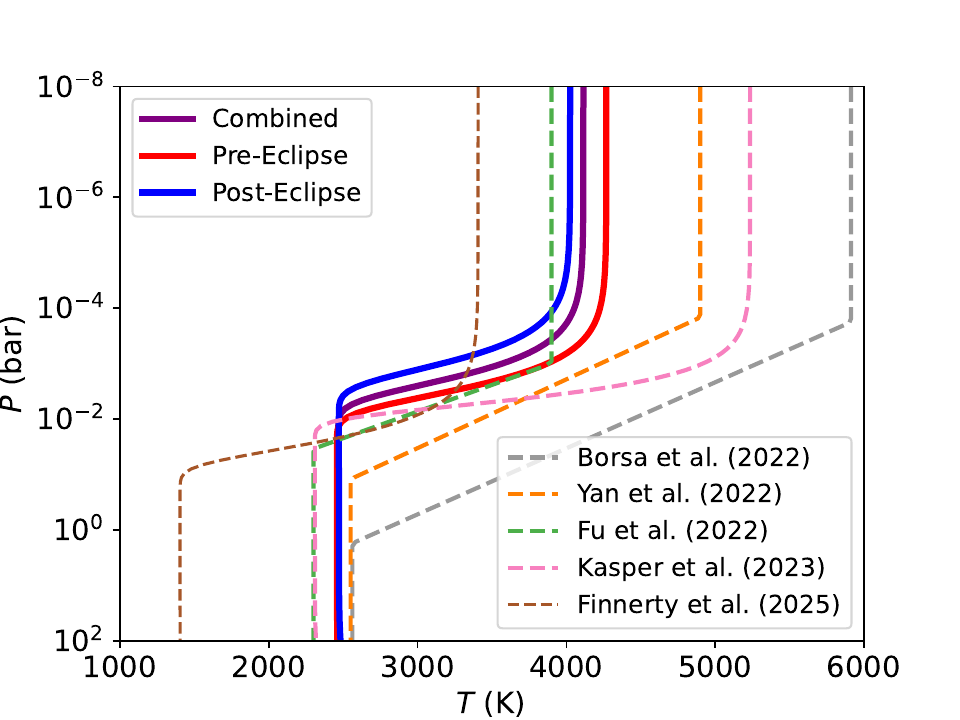}
\caption{Comparison of our median retrieved P-T profiles for our pre-eclipse, post-eclipse, and combined datasets along with the published P-T profiles retrieved by \cite{borsa+22}, \cite{Yan+22}, \cite{Fu+22}, \cite{Kasper+23}, and \cite{Finnerty+25}. Each of these published P-T profiles is retrieved from datasets including both pre- and post-eclipse observations, except for \cite{Finnerty+25}, who only observed KELT-20 b post-eclipse.} We note that we used only an approximation to the P-T profile retrieved by \cite{Fu+22} as described in \cite{Johnson+23}. 
\label{fig:comparison}
\end{figure*}

\begin{figure*}[t]
\centering
\includegraphics[width=\textwidth]{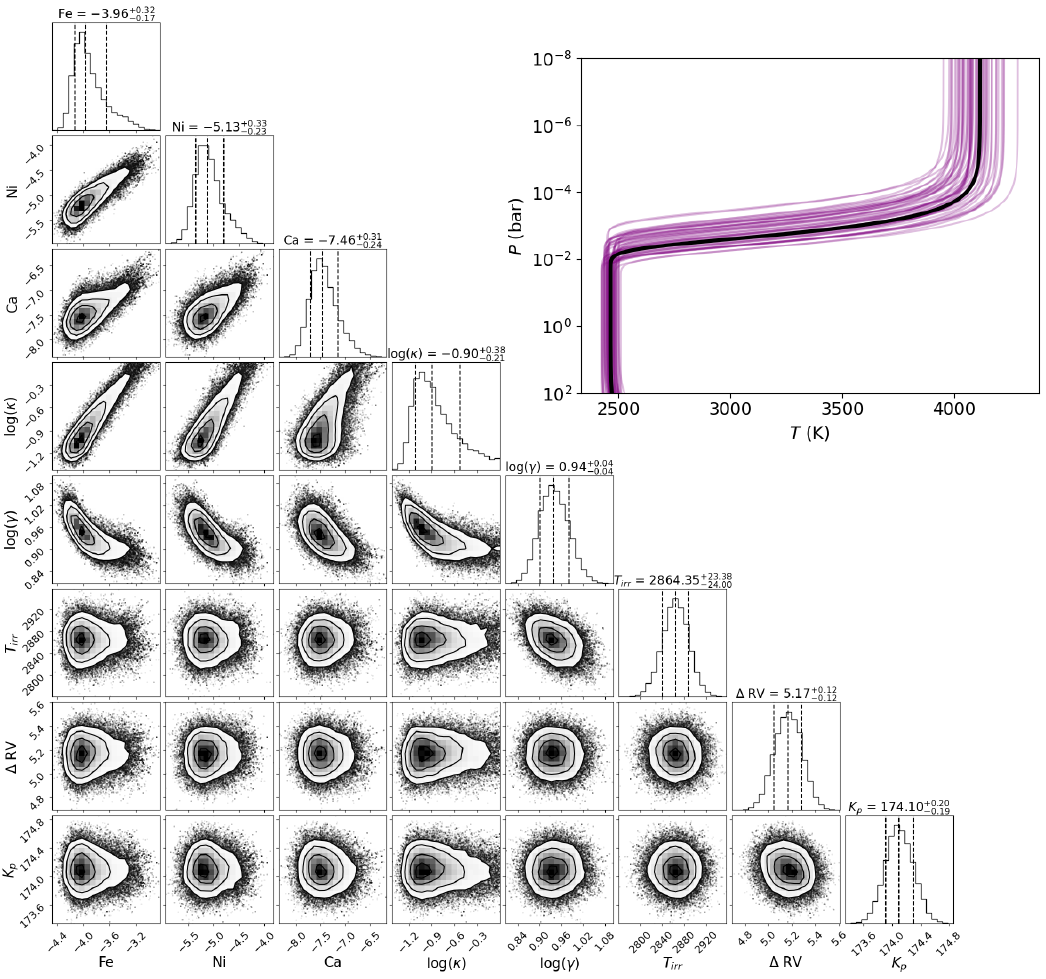}
\caption{Full retrieval results for KELT-20 b. The medians and 1$\sigma$ confidence intervals are denoted in
the histograms with dashed lines and are listed above each histogram panel, as well as in Table \ref{tab:parameters}. \textit{Upper right}: The median retrieved P-T profile shown as a thick black line, with thinner purple lines representing 50 P-T profiles randomly extracted from the MCMC posteriors.}
\label{fig:combined}
\end{figure*}

\begin{figure*}[t]
\centering
\includegraphics[width=\textwidth]{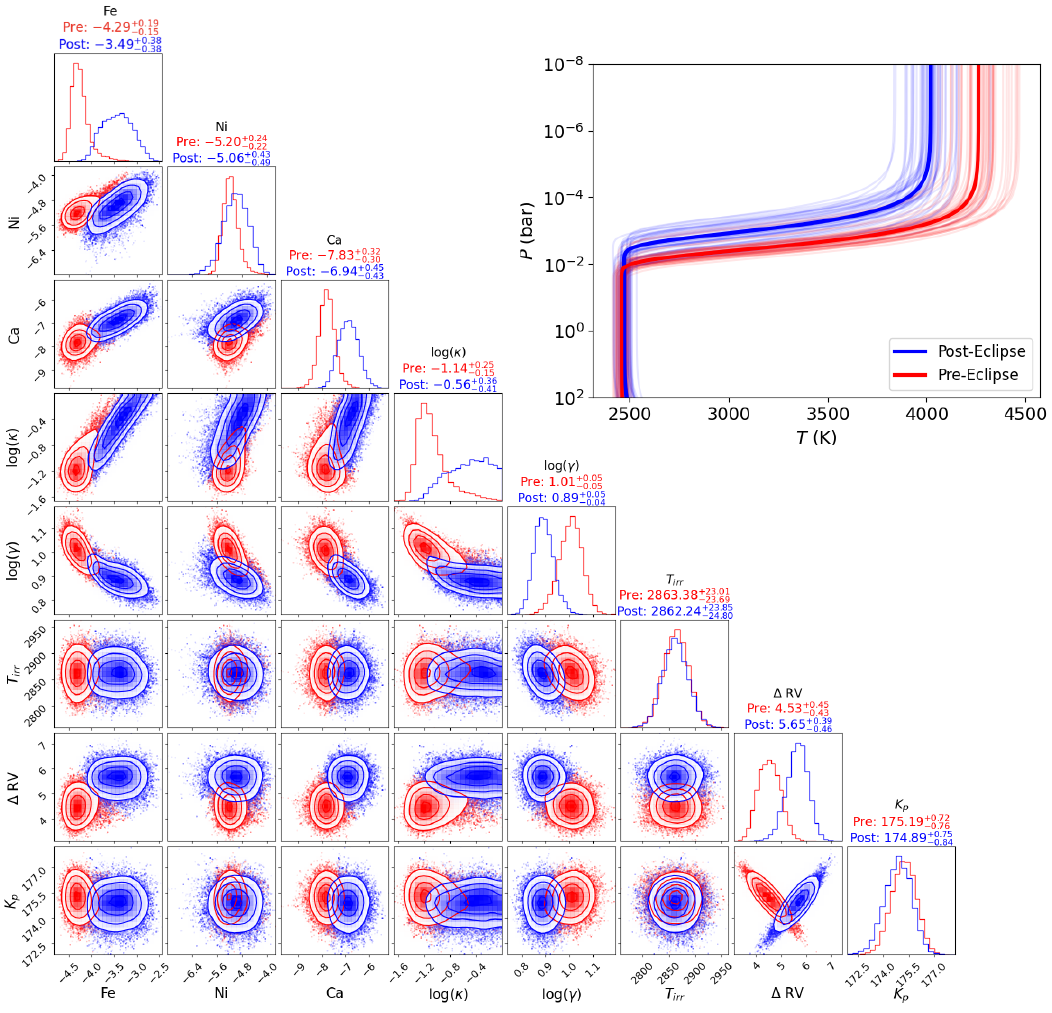}
\caption{Same as Figure \ref{fig:combined}, but with the pre-eclipse (red) and post-eclipse (blue) datasets shown separately.}
\label{fig:phase}
\end{figure*}

While there is a broad spread in the previously reported P-T profiles for KELT-20 b \citep[e.g.][]{Finnerty+25, Kasper+23, borsa+22, Yan+22, Fu+22}, our retrieved P-T profiles for the combined, pre-, and post-eclipse datasets lie within the scatter of published results, as shown in Figure \ref{fig:comparison}. The reason for this wide spread in retrieved P-T profiles is unclear, but like the abundance disagreements discussed in \S\ref{sec:abundances}, this could be due to either real heterogeneity of the atmosphere, or differing assumptions in the retrieval frameworks. 

We constrain difference in upper temperature between the post-eclipse and pre-eclipse datasets on the order of a few hundred kelvin.  GCMs across a wide range of giant planets predict a prominent east-west flow from equatorial jets in the deep atmosphere that weaken at higher altitudes, where the day-night flow begins to dominant \citep[e.g.][]{Showman+Guillot, Kataria+16, Flowers+19}. 
For example, \cite{Wardenier+25}, Figure 2 shows drag free models for WASP-76 b for two different pressure levels ($P=10^{-1}$ bar and $P=10^{-4}$ bar). In the $P=10^{-1}$ bar model, the pre-eclipse (phase = 0.375) dayside temperature is significantly hotter than the post-eclipse (phase = 0.625) dayside. However, in the $P=10^{-4}$ bar models the temperatures on the pre- and post-eclipse phases are about the same. Therefore these data probe a pressure regime where day-night winds should begin to dominate over the equatorial jet. These results suggest that a minimal temperature gradient between KELT-20 b's morning and evening sides persists in this pressure regime, and could be due to a weakened east-west wind flow dumping hot atmospheric material on the dayside. 

\section{Conclusions}\label{sec:conclusion}
We have presented five high-resolution emission spectroscopy observations of the UHJ KELT-20 b using the PEPSI spectrograph, covering the phases before and after secondary eclipse. We applied a Bayesian retrieval framework to constrain the VMRs of Fe, Ni, and Ca, providing the first constraints reported on Ni and Ca in emission for KELT-20 b. We applied the same retrieval framework to the pre-eclipse datasets, probing the evening side, and to the post-eclipse datasets, probing the morning side, and find a slightly higher abundance of the detected refractory species on the morning side.

We compare our results to a FastChem equilibrium chemistry model and find that our retrieved abundances suggest a super-solar composition, ranging between 10-30 $\times$ solar. However, we caution against drawing strong conclusions from our absolute abundance constraints due to the degeneracies between the absolute abundances and the opacity level. Instead, we rely on the robustness of abundance ratios to further interpret our results. We calculate the nickel-to-iron and calcium-to-iron abundances for the pre-eclipse, post-eclipse, and combined posteriors.
We find these ratios to all be consistent with each other within 1$\sigma$. This constraint, when combined with the [Ni/Fe] and [Ca/Fe] values calculated in \cite{Gandhi+23} during transit, suggests that even though the abundances of the neutral species vary significantly during multiple phases of KELT-20 b's orbit, their abundance ratios remain relatively constant. This result underscores the robustness of constraining abundance ratios with high-resolution spectroscopy, as previously demonstrated in other HRS studies \citep[e.g.][]{Gibson+22, Maguire+24, Ramkumar+25}

Our measured pressure-temperature profiles fall within the spread of previously published results. 
We constrain a slightly higher temperature at the top of the thermal inversion on the evening side as compared to the morning side, suggesting possibly longitudinal temperature differences across the dayside of KELT-20 b. This analysis demonstrates that additional information can be gained when performing retrievals on multiple phases of high-resolution spectra. 

Performing retrievals on separate orbital phases can help us probe the longitudinal patterns and the altitudinal temperature structure, allowing us to study these planets as 3D structures. While our study employs a particularly high SNR, large phase range dataset, these methods can be applied to other datasets in both emission and transmission with smaller phase ranges. 
Further phase-resolved atmospheric studies will continue to build our understanding of the evolving dynamics and chemical processes of these planetary systems.

\section{Acknowledgments}
\noindent The authors thank the anonymous referee for the thorough and insightful report that greatly improved this work.

MCJ, JW, and JK are supported by NASA Grant 80NSSC23K0692.

This research was supported in part by the University of Pittsburgh Center for Research Computing through the resources 
provided. Specifically, this work used the HTC cluster, which is supported by NIH award number S10OD028483.

The LBT is an international collaboration among institutions in the United States, Italy and Germany. LBT Corporation Members are: The Ohio State University, representing OSU, University of Notre Dame, University of Minnesota and University of Virginia; LBT Beteiligungsgesellschaft, Germany, representing the Max-Planck Society, The Leibniz Institute for Astrophysics Potsdam, and Heidelberg University; The University of Arizona on behalf of the Arizona Board of Regents; and the Istituto Nazionale di Astrofisica, Italy. Observations have benefited from the use of ALTA Center (\url{alta.arcetri.inaf.it}) forecasts performed with the Astro-Meso-Nh model. Initialization data of the ALTA automatic forecast system come from the General Circulation Model (HRES) of the European Centre for Medium Range Weather Forecasts.

\bibliography{bib}{}
\bibliographystyle{aasjournal}

\restartappendixnumbering
\clearpage

\end{document}